\documentclass[10pt]{article}
\usepackage{amsmath, amsthm, amssymb}
\usepackage[latin1]{inputenc}
\usepackage{multirow}
\usepackage[english]{babel}
\usepackage{relsize}
\usepackage[pdftex]{graphicx}

\usepackage{enumerate}
\usepackage{subfig}
\usepackage{verbatim}  

\begin{document}
\begin{titlepage}
\vskip0.5cm
\vskip0.5cm
\begin{center}
{\Large\bf  Ising model description of Long Range correlations in DNA sequences.}
\end{center}
\vskip1.3cm
\centerline{A. Colliva$^{a}$, R. Pellegrini$^{b}$, A. Testori$^{a}$, M. Caselle$^{a}$}
 \vskip1.0cm
 \centerline{\sl  $^a$ Dipartimento di Fisica dell'Universit\`a di Torino and I.N.F.N. sez. di Torino,}
 \centerline{\sl Via Pietro~Giuria 1, I-10125 Torino, Italy}
 \vskip0.3 cm
 \centerline{\sl  $^b$ Physics Department, Swansea University, Singleton Park, Swansea SA2 8PP, UK}
 \vskip0.4 cm
 \centerline{\sl
e--mail: \hskip 1cm
 colliva(ropelleg)(testori)(caselle)@to.infn.it}
 \vskip1.0cm
\begin{abstract}
We model long range correlations of nucleotides in the human DNA sequence using the long range one dimensional Ising model. We show that for distances 
between $10^3$ and $10^6$ bp the correlations show a universal behaviour and may be described by the non-mean field limit 
of the long range 1d Ising model. This allows us to make some testable hypothesis
on the nature of the interaction between distant portions of the DNA chain which led to the DNA structure
that we observe today in higher eukaryotes. 

\end{abstract}
\end{titlepage}

\section{Introduction}

One of the most surprising features of higher eukaryotes genomes is the presence of long range correlations in the composition of the DNA sequence.
These correlations were discovered more than 20 years ago \cite{Peng92} when the first long continuous DNA sequences became available.
Soon after this discovery several evolutionary models were proposed \cite{li:PhysRevA.43.5240,messer_solvable_2005,Grosberg93,Buldyrev93,Goodsell94,
Vaillant03,Audit01,Audit02,Vaillant05, Koroteev11,li_universal_2005,li_spectral_2004} 
to explain this behaviour and compared with the growing collection of genomic data.

Thanks to next generation sequencing projects an impressive amount of whole-genome sequences is now available, and the composition of genomic DNA
can be studied systematically over a wide range of scales and organisms. This makes it now possible to assess the various models proposed for the description of these
long-range correlations in a more extensive way. The statistical analysis is quite 
intricate since genomic DNA is a rather ``patchy'' statistical
environment: it consists of genes, non-coding regions, repetitive elements etc. 
Despite this complexity a few general 
results are by now well established. 
\begin{itemize}
\item
These correlations extend over a range much longer than previously expected and reach distances of the order of $10^7$ bp
(see fig.\ref{pic:ens}).
\item
They show a power law behaviour, with exponents characterised by a remarkable degree of universality, with 
very small variations across the human chromosomes \cite{li_universal_2005} and between the human and mouse genomes 
\cite{li_spectral_2004}.
\item
In the human case these exponents take values in the range [0.05-0.30]. These values are 
much smaller than usual scaling exponents observed in genomic data. For instance they are much smaller than those of
chromatine contacts within chromosomes as extracted from Hi-C experiments (we shall discuss in more detail this issue in sect.4). 
\end{itemize}

These features pose severe constraints on the models proposed to explain the correlations. 
In particular their universality and the unusual long range scales 
are features typical of critical systems
and suggest a modelling strategy characterised by scale invariance and 
universality, i.e. models which do not depend too much 
on the microscopic details and on genomic features with a fixed reference scale.
  
A very interesting model which fulfils these conditions is the so called ``expansion-randomisation" (E-R) model 
proposed by Li in 1989 \cite{li:PhysRevA.43.5240} and solved exactly in \cite{messer_solvable_2005}. The stationary state of the model 
is characterised by the expected long range correlations and is largely independent from microscopic details \cite{messer_solvable_2005}.
The only problem of this model is that in order to match the observed exponents it requires duplication and insertion rates
  much larger than the one derived from the actual expansion rate of our genome. Indeed by comparing the human sequence with that
   of other mammals (and in particular with the mouse genome) one can see that in the last 100 Myr, i.e. since the mammalian radiation,
    the human genome was almost stable or at most expanded very slowly and that its expansion was mainly due to retrotransposons insertions. 
    Thus it is likely that the E-R model should be complemented with some other evolutionary process able to enforce long range correlations 
    without requiring an expanding genome. In particular the unusual range of these correlations suggests the introduction of non-local interactions in the evolutionary process.  
    
Following this line of reasoning in this paper we propose an evolutionary model based on non-local moves
which - as the  E-R one - is able to reproduce the large distance correlations observed in the DNA sequences 
 but does not require an expanding genome.
Notwithstanding the intrinsic complexity of the non-local interactions, several features of the model, and in particular the scaling exponents, can be predicted very accurately because the stationary state of the model
 can be mapped  into the equilibrium state of a (very peculiar) statistical model, the so called ``long range one dimensional Ising model", for which several exact and approximate results exist. In particular, differently from the ordinary (short range) Ising model, this model admits 
a critical point also in one dimension and in the neighbourhood of this point displays long range correlations exactly of the type observed in the human DNA sequence. 

As we shall discuss below, we think that our model and the E-R one should be considered 
as  complementary processes which were probably both active in the evolutionary path of higher eukaryotes and both contributed to shape the long range features of the genome that we observe today.  

This paper is organised as follows. In the next section we shall discuss the statistical analysis of the DNA sequences and recover 
the large distance correlations mentioned above. In the third section we shall propose our model, map it into the 1d long range 
Ising model and discuss its main properties. The fourth section is devoted to a tentative biological interpretation of our results while the last one is devoted 
to a few concluding remarks.

\section{Sequence Analysis}

\subsection{DNA correlators}
We computed the base-base correlation 
function along the lines discussed for instance in \cite{li_study_1997, messer_corgen--measuring_2006}.

We defined a map from the 4 letter alphabet to a binary set as follows
\begin{equation} \label{eq:map}
\begin{cases}
 \{A,T\}\rightarrow {-}
 \\[.9em]
 \{C,G\}\rightarrow {+}
\end{cases}
\end{equation}
We shall denote in the following these pairs of nucleotides as ``spins"  $\sigma=\pm 1$ .
This identification greatly simplifies the analysis while keeping the full complexity
of long range correlations of the genome and was adopted also in previous analyses of these correlations \cite{messer_solvable_2005}. 

\begin{figure}[h]
\begin{center}
\hspace*{-1.0cm}
{\includegraphics[width=.53\columnwidth]{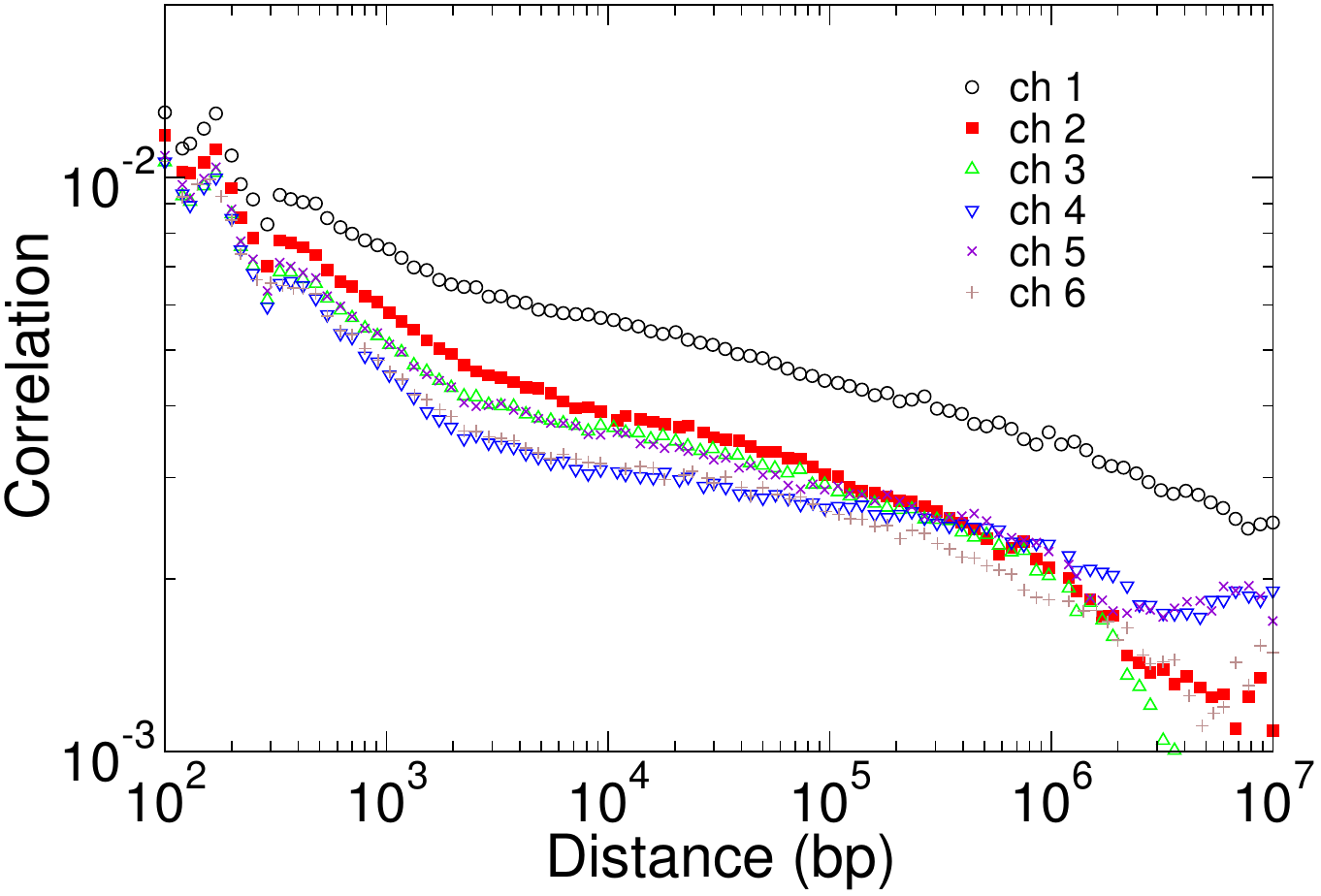}} 
{\includegraphics[width=.53\columnwidth]{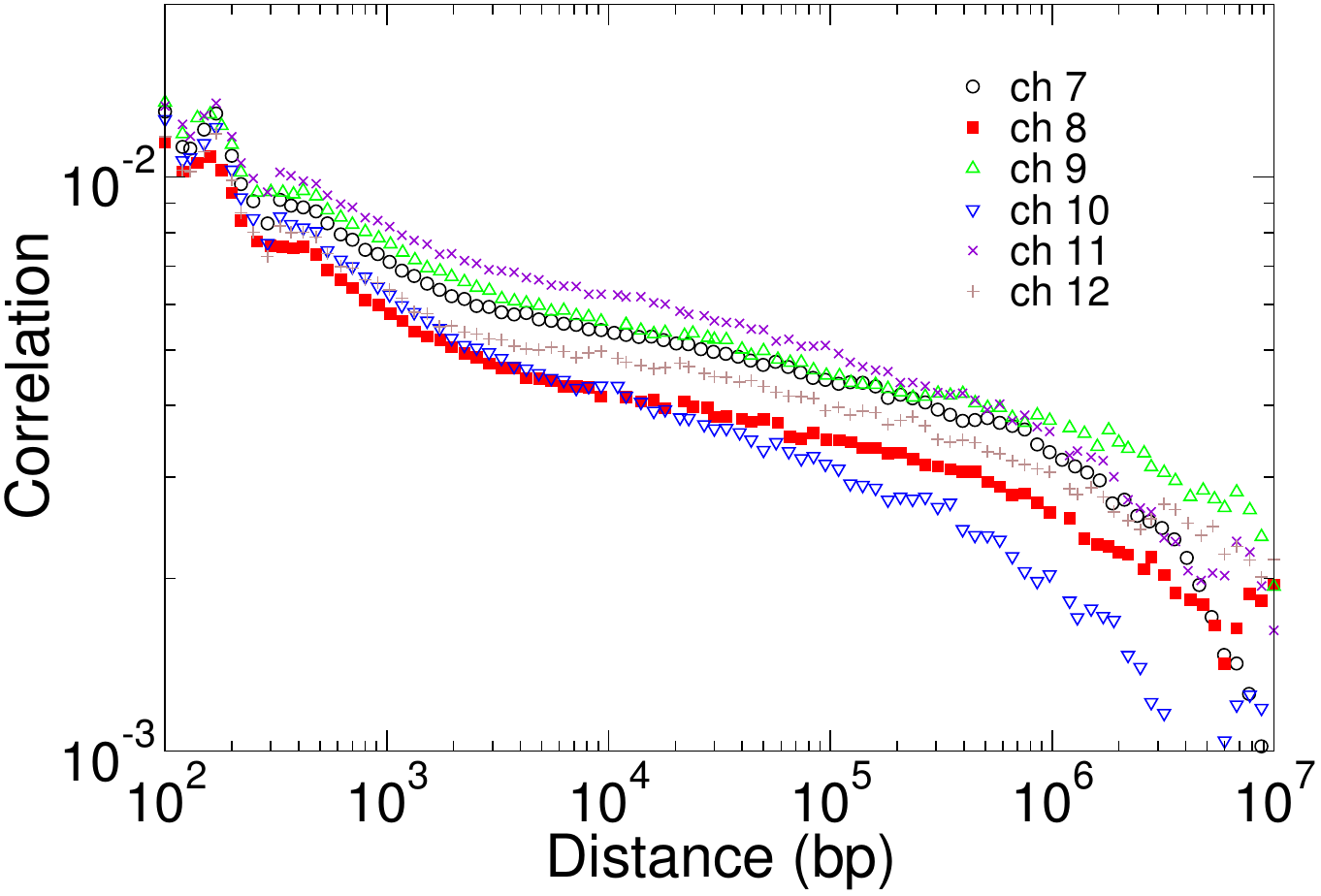}} \\
\hspace*{-1.0cm}
{\includegraphics[width=.53\columnwidth]{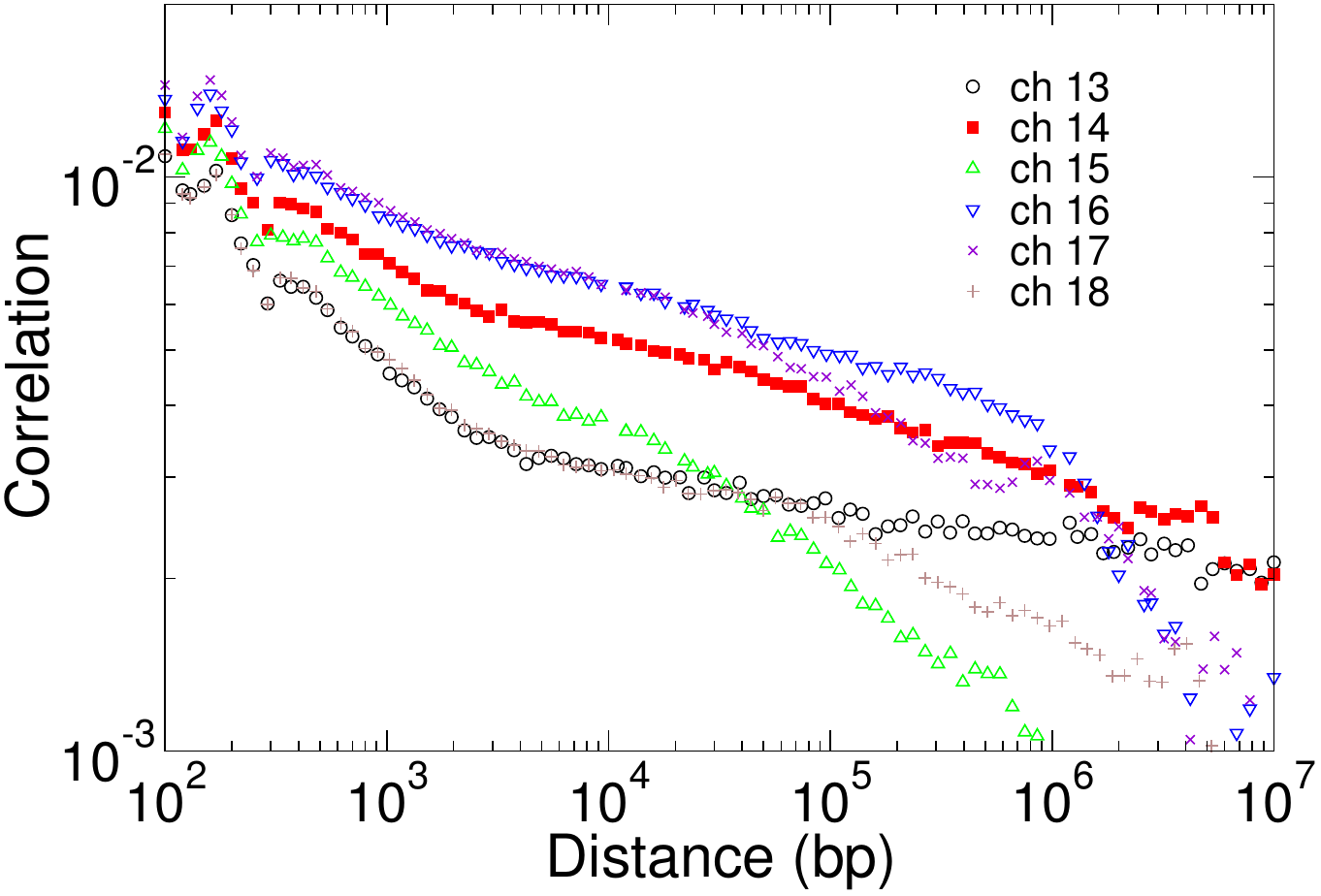}} 
{\includegraphics[width=.53\columnwidth]{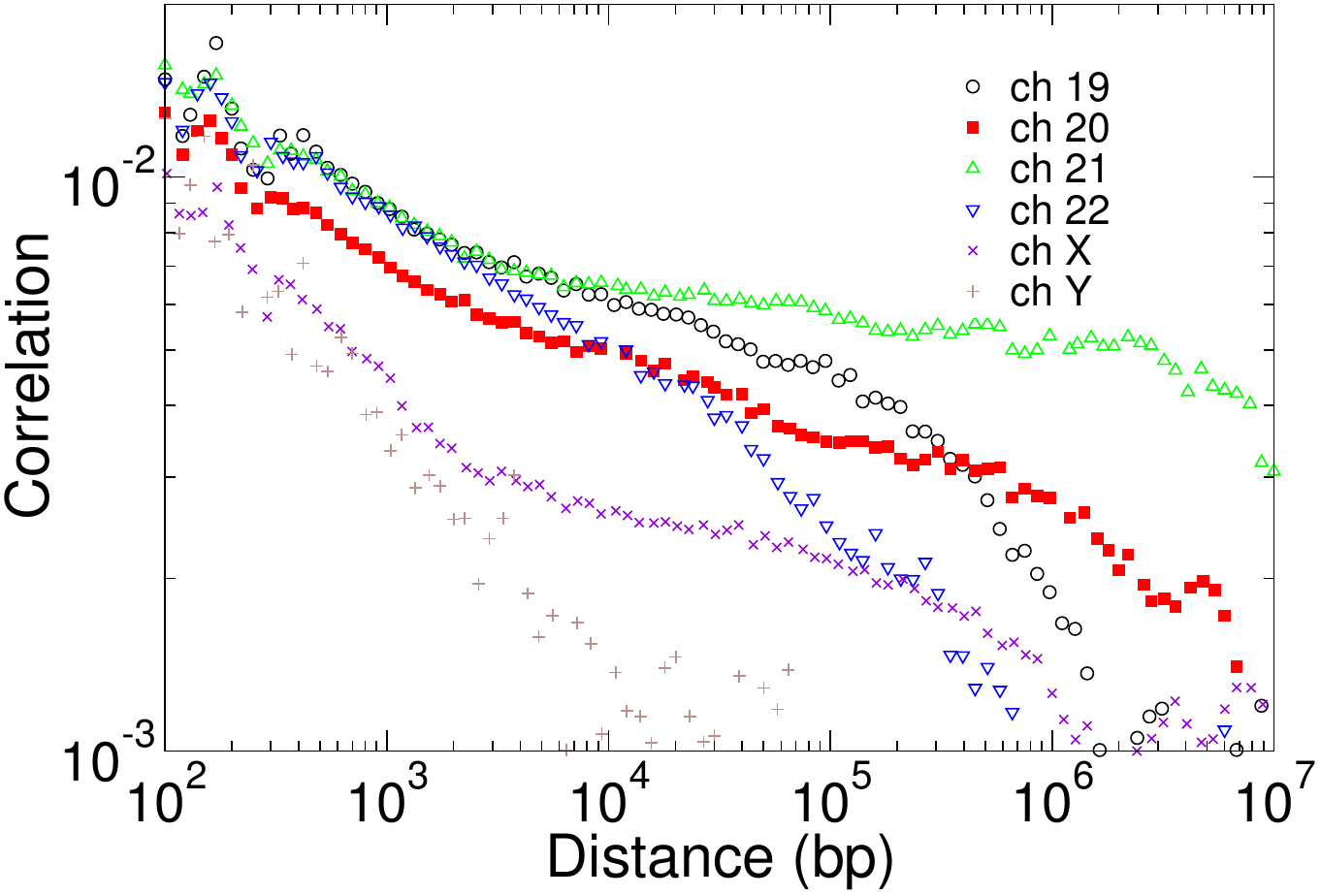}}
\end{center}
\caption{Color online. The combination $\Gamma_{++}(d)+\Gamma_{--}(d)$ for the human chromosomes in a log-log scale. 
Notice the similarity among the different curves and the wide range of validity of the power law behaviour.}
\label{pic:ens}
\end{figure}

We computed the  correlation function at a given distance $d$ using a frequency-count estimator \cite{li_study_1997, messer_corgen--measuring_2006}.
\begin{equation}
{\Gamma_{\alpha\beta}(d)}=\frac{N_{\alpha\beta}(d)}{N}-\frac{N_\alpha}{N}\frac{N_\beta}{N} \qquad \alpha, \beta\in\{+,-\}
\end{equation}
where $N$ is the total length of the sequence, $N_{\alpha\beta}(d)$ is the number of occurrences of $\alpha$ and $\beta$ at distance $d$ and 
$N_\alpha$ denotes the total number of spins of type $\alpha$.
Given the symmetries of the system it is enough for our purpose to 
compute only the positive correlators in which $\alpha= \beta$.
The curves that we obtained for the combination $\Gamma_{++}(d)+\Gamma_{--}(d)$ are plotted in log-log scale in figure \ref{pic:ens} for various human chromosomes.

Looking at the figures it is easy to identify three regimes. A short range regime, below 1 kilobase (kb), which is dominated by the fine structure
 of the sequence (regulatory regions, correlations induced by nucleosomes, codon bias...), an intermediate regime 
between $2 \times 10^3$ and $10^5$ (or $10^6$ for the longest chromosomes) base pairs  (bp) where the slopes of the correlators show a 
remarkable degree of similarity among different chromosomes and a rather clear power law behaviour of the type
\begin{equation}
{\Gamma_{++}(d)}\sim d^{-\gamma}
\label{defgamma}
\end{equation}
 can be identified, and a large distance region for $d > 10^6$ 
bases in which the correlation function drops drastically and no evidence of a universal behaviour can be found. In the following we shall concentrate on the intermediate region. Our goal will be to construct an evolutionary model able to reproduce the observed 
power law correlators.

It is natural to identify these three regimes with those which are typically observed in correlators of standard statistical mechanics models in the vicinity of a critical point  (think for instance to the 2d Ising model as an example): a short range regime which is dominated by ``lattice artefacts" and depends on the precise microscopic definition of the model, a large distance regime
for distances larger than the correlation length whose behaviour is dominated by the spectrum of the theory in which the correlation function decreases exponentially, and an intermediate ``universal" regime in which the correlation function decreases with a  power law and is dominated by the nearby critical point (and for this reason is universal, i.e. only depends on the universality class of the critical point).

The only non trivial point of this identification is that as it is well known no critical behaviour (i.e. no long range correlations) may exist in one 
dimensional equilibrium statistical models with short range interactions. This is the first indication that we shall have to consider in our analysis one dimensional models with long range interactions. We shall come back to this point in the next section.

\subsection{Power law fitting}

We fitted ${\Gamma_{++}(d)}$ with equation (\ref{defgamma}) for all the chromosomes. In order to test the stability of the results with respect to the range of
distances included in the fits
we perfomed two different sets of fits, choosing first a conservative range $2\times10^3<d< 10^5$ and then a much larger window $5\times10^2<d< 10^6$.

\begin{figure}[h] 
 \begin{center}
 \hspace*{-1.cm}
  \includegraphics[width= 1.\textwidth]{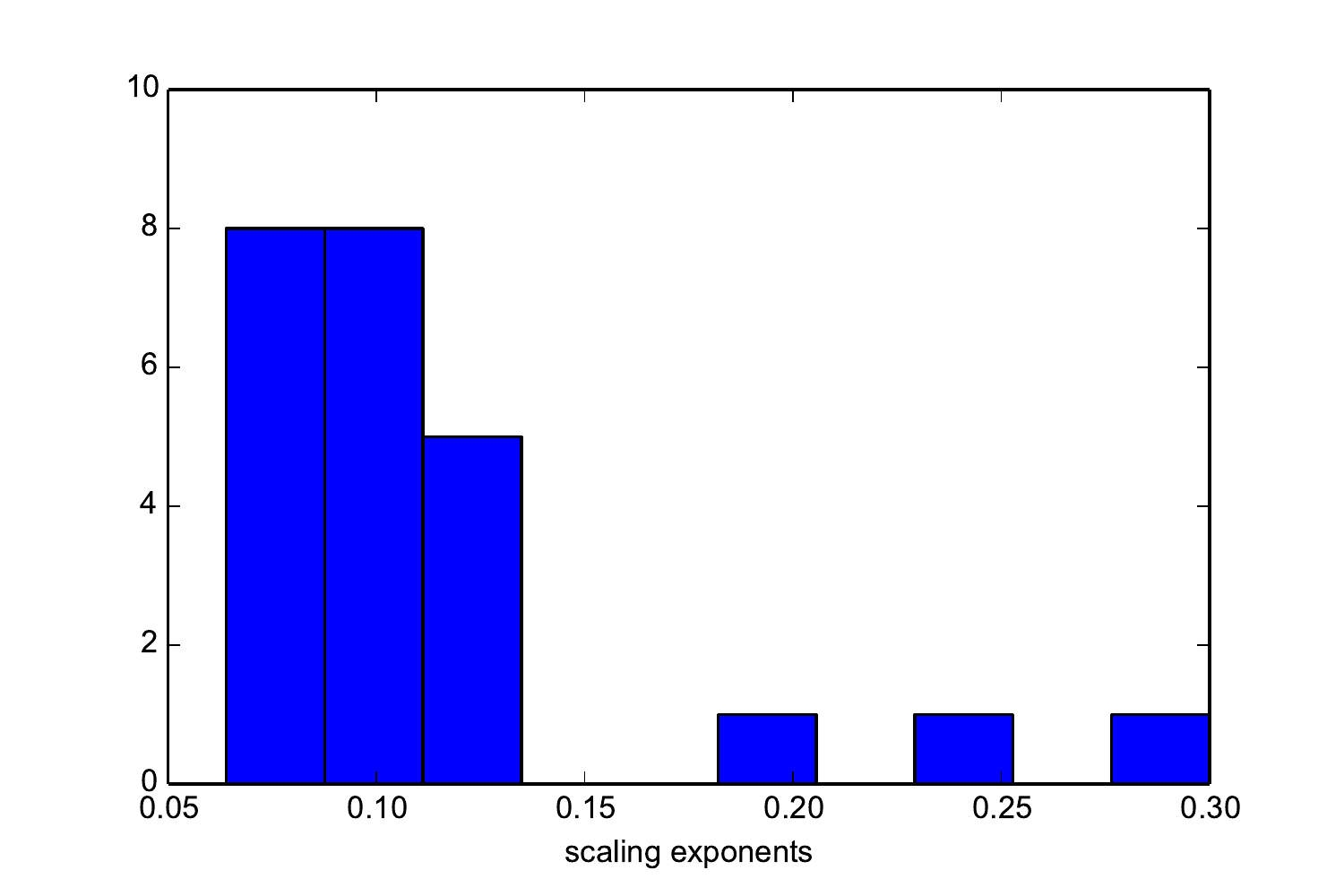}
  \end{center}
  \caption{\footnotesize{Color online. Distribution of the  scaling exponents $\gamma$.}} \label{pic:hist}
\end{figure}

The values of the scaling exponents extracted from the fits are reported in tab. \ref{tab:gamma} and, for the more conservative choice of distances in 
fig. (\ref{pic:hist}). Comparing the two 
columns of tab.\ref{tab:gamma} one may obtain a rough estimate of the systematic uncertainties induced by the fitting range.

\begin{table}[h]
\begin{tabular}{c c c}
\hline
\textbf{chromosome} & \textbf{$\gamma_1$} & \textbf{$\gamma_2$} \\ \hline
1 & 0.0948(2) & 0.1045(2) \\
2 & 0.1158(7) & 0.1353(5) \\
3 & 0.0901(3) & 0.1137(4) \\
4 & 0.0754(8) & 0.0825(8) \\
5 & 0.0940(6)& 0.1060(6) \\
6 & 0.0840(11) & 0.1139(10) \\
7 & 0.0786(2) & 0.0950(3) \\
8 & 0.0876(9)& 0.1014(7) \\
9 & 0.0919(7)& 0.0964(6) \\
10 & 0.1226(7) & 0.1524(5) \\
11 & 0.0929(4) & 0.1109(2) \\
12 & 0.0724(5) & 0.0958 (4)\\
13 & 0.0841(11) & 0.0822(9) \\
14 & 0.0956(6) & 0.1160(3) \\
15 & 0.1899(18) & 0.2451(9) \\
16 & 0.1040(8) & 0.1192 (4)\\
17 & 0.1291(12) & 0.1638(6) \\
18 & 0.0874(5) & 0.1315(7) \\
19 & 0.1276(3)& 0.1720(8) \\
20 & 0.1317(11) & 0.1324(6) \\
21 & 0.0640(10)& 0.0658(6) \\
22 & 0.2499(17) & 0.2868(15) \\
 X & 0.0965(1) & 0.1650(3) \\
 Y & 0.2970(16) & 0.3650(8) \\
\hline
\end{tabular}
\caption{Values of the scaling exponents extracted from the fits. The first set of values, denoted by $\gamma_1$ correspond to the more conservative 
range of distances $2\times10^3<d< 10^5$, while the second set of data, denoted by $\gamma_2$, corresponds to the larger range of fitted distances 
$5\times10^2<d< 10^6$. The statistical errors of the fits are reported in parenthesis while the comparison between the two 
columns gives an idea of the systematic uncertainties induced by the fitting range.}
\label{tab:gamma}
\end{table}

It is interesting to notice that, with the exception of chromosomes 15, 22 and Y, the values of $\gamma$ obtained with the more conservative choice of fitted distances 
are all contained in a rather narrow interval $[0.06-0.13]$. With the other choice of distances the range of values of $\gamma$ increases, but it remains confined within the
range $[0.06-0.17]$.
The non-trivial behaviour of chromsomes 15, 22 and Y had been already noticed and discussed in \cite{li_universal_2005}.

\subsection{Stability of the power law behaviour under sequence coarsening.}
\begin{figure}[h] 
 \begin{center}
 \hspace*{-1.5cm}
  \includegraphics[width= 1.2\textwidth]{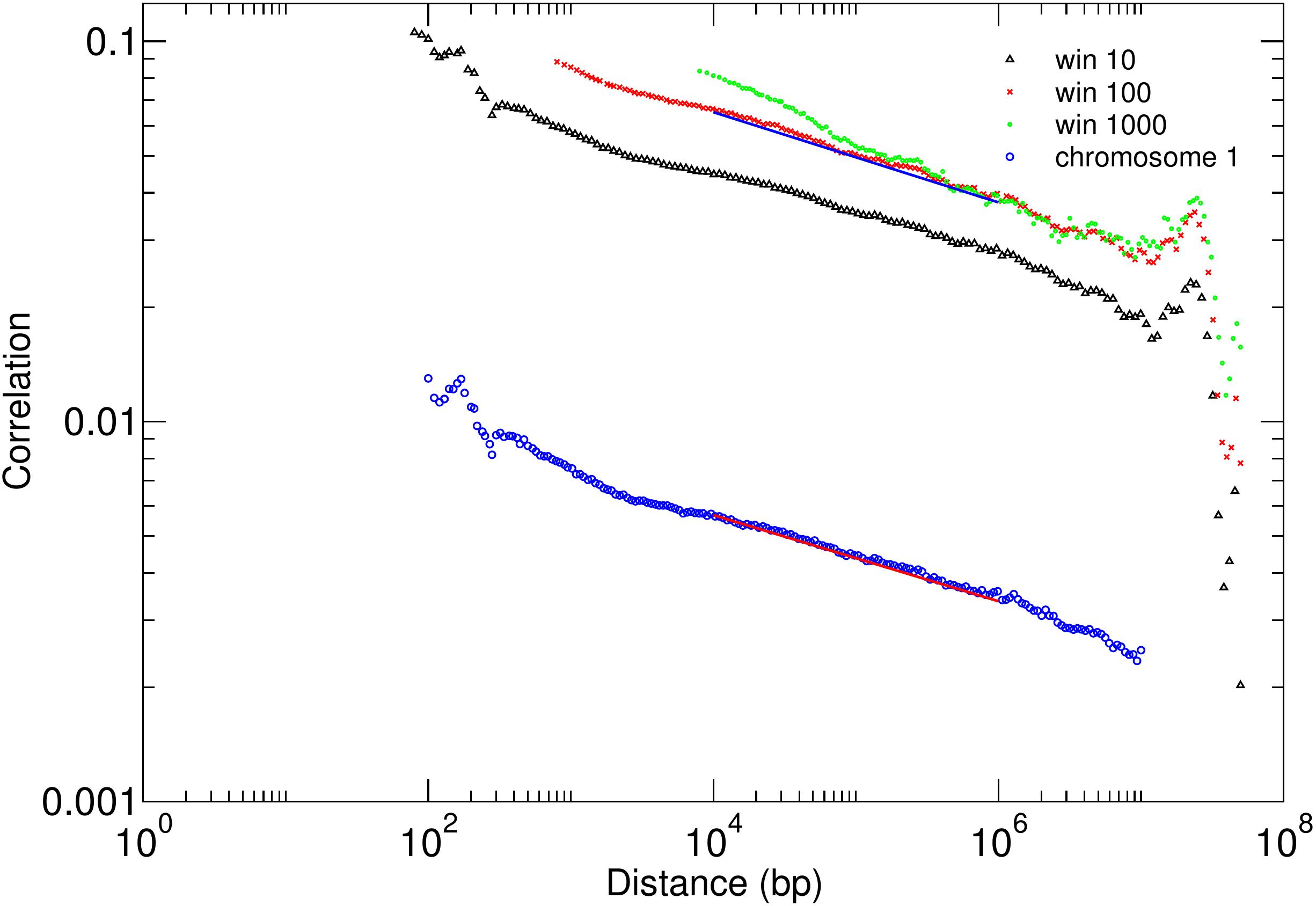}
  \end{center}
  \caption{\footnotesize{Color online. Comparison of base-base correlations  in chromosome 1 (lower part of the figure, blue circles),  with their renormalized version (in the upper part of the
  figure) obtained with a renormalization step of 10 bp (black triangles), 100 bp (red crosses) and 1000 bp (green dots).}} \label{pic:renorm}
\end{figure}

A very interesting and non trivial feature of the DNA correlations that we are studying is that their power law behaviour is very stable against renormalization.
We implement the renormalization transformation using a simple majority rule,
coarse-graining the sequence and substituting each window with a sign chosen with a majority rule.
The result of this process is represented in the case of chromosome 1 in the upper part of the graph in figures \ref{pic:renorm} for various sizes of the renormalization
window. All the other chromosomes show essentially the same behaviour. 
It is easy to see that up to window sizes of 100 bp nothing changes and that only for window sizes of 1000 bp one can observe some finite size effect at
short scale which disappears at larger distances where the original exponent of the power law decay is recovered also in this case. 
From a statistical mechanics point of view, this remarkable stability tells us that the original sequence is already very near to a critical point and that,
irrelevant operators (i.e. subleading exponents), if present, should have an almost negligible coupling. From a biological point of view this is telling us that, more than
the single nucleotide, the basic sequence element driving the observed correlations are sequences of intermediate length (from a few tens up to a few hundreds of bases) 
with a small AT or CG bias. We shall come back to this point in the last part of the paper.

\section{The model}
The model that we propose is very simple. It is defined on a one dimensional lattice. At each time step
\begin{enumerate}
 \item  Randomly choose a spin $\sigma_i$ (i.e. a nucleotide) from the lattice.
 \item Randomly choose a second spin $\sigma_j$ and fix its sign to be the same of $\sigma_i$ with probability $p_+$ ($p_-$) if $\sigma_i=+1$ ($\sigma_i=-1$) defined as 
 \begin{equation}
  p_{\pm} = \frac{e^{\beta |i-j|^{-\alpha}\pm h}}{e^{\beta |i-j|^{-\alpha}\pm h}+e^{-(\beta |i-j|^{-\alpha}\pm h)}}  
  \label{heatbath}
 \end{equation}
  where $|i-j|$ denotes the distance between $i$ and $j$ 
\end{enumerate}
 $\beta,\alpha$ and $h$ are parameters which we shall discuss 
  later but they will be always such that we may safely approximate the probability as 
 \begin{equation}
  \label{prob}
  p_{\pm} \sim  \frac12 + \frac{\beta}{|i-j|^{\alpha}}\pm h
 \end{equation}
 i.e the sum of a pure random choice with a drift 
 plus an excess probability  
 $p \sim  \frac{\beta}{|i-j|^{\alpha}}$ to align the two spins between them.

These two steps define a Markov chain  
which has as stationary state the probability distribution of the 1d long range Ising model defined by the following Hamiltonian:
\begin{equation} \label{ham}
 \mathcal{H}=\mathsmaller - \sum_{x y} J (x-y)^{-\alpha} \sigma_x \sigma_y - h \sum_x \sigma_x.
\end{equation}
More precisely, the above steps define one of the possible choices for a  Montecarlo algorithm which 
simulates this particular model\footnote{In particular, the choice of eq. (\ref{heatbath}) defines the so called ``heat-bath" algorithm. Notice, as a side remark, 
that if one is actually interested in simulating the model, the heat bath is not the best option for a non-local model like this one and that 
cluster based models like the
one discussed in \cite{Luijten02} are much more efficient.}.

This model is very interesting since it is the simplest example of a one dimensional spin model with a critical point and
 has been the subject of considerable theoretical efforts in the last 50 years. Its phase diagram is rather complex and
depends on the parameter $\alpha$ which must be greater than one to have a well defined finite expression for the interaction energy. 
As $\alpha$ increases  the model is characterised by three different behaviours.
\begin{itemize}
\item  For $1 < \alpha < 1.5$ the model admits a second order phase transition for  $h=0$ and for a critical value $\beta_c(\alpha)$ which depends on the precise value of
$\alpha$. For $\beta > \beta_c$ the $Z_2$ symmetry of the model is spontaneously broken and the system is characterised by a non-zero magnetisation. Using standard
renormalization group analysis \cite{Fisher72} it can be shown that   
the critical point belongs to the mean field universality class.
\item For $1.5 < \alpha < 2$ the model still has a second order phase transition, but the universality class is not any more of the mean field type. The critical
exponents vary as functions of $\alpha$.
\item
For $\alpha > 2$ the system behaves as a short range Ising model and since the lattice is one dimensional, there is no more a phase transition and the $Z_2$ symmetry is
unbroken for any finite value of $\beta$.
\end{itemize}

The most important result for the scope of the present paper is that, due to the continuous nature of the phase transition,
in the range $1 < \alpha < 2$ in the vicinity of the critical point we expect long range correlations between spins. These correlations are controlled by the scaling
dimension $d_\phi$ of the spin operator
\begin{equation}
<\sigma_i \sigma_j> \sim \frac{1}{|i-j|^{2d_\phi}}.
\end{equation}
The scaling dimension $d_\phi$ depends on $\alpha$. In the mean field regime, where it can be evaluated analytically, the relation is very simple: $d_\phi= 1-\alpha/2$,
which implies
\begin{equation}
\gamma\equiv 2d_\phi=2-\alpha
\label{rel}
\end{equation}
 and
leads to values of $\gamma$ in the range $1 > \gamma > 0.5$, i.e. larger than those which we have seen are typical of genomic correlators. Thus we are bound to study
the model in the non-mean field regime.
Very few results are known exactly outside the mean field regime but, remarkably enough, it can be shown using the $\epsilon$ expansion 
that $d_\phi$ is not renormalized up to the third order. Indeed it is commonly believed that it should keep its mean field value 
to all orders in the $\epsilon$ expansion\footnote{Notice that this result holds only in the one dimensional case. In more than one dimension it can be shown that it holds
only up to the value of $\alpha$ for which $d_\phi$ reaches the values it has in the corresponding short range Ising model. 
See \cite{Angelini14} for an updated review of these
results.}. This is exactly what we need to fix the value of $\alpha$ in our model. In order to match the observed genomic
correlations $\alpha$ should range in the region $1.9 > \alpha > 1.7$ which gives for the anomalous dimensions values in the range $0.3 > \gamma > 0.1$ where most of the
genomic correlators lie.

\begin{figure}[h] 
 \begin{center}
 \hspace*{-1.5cm}
  \includegraphics[width= 1.2\textwidth]{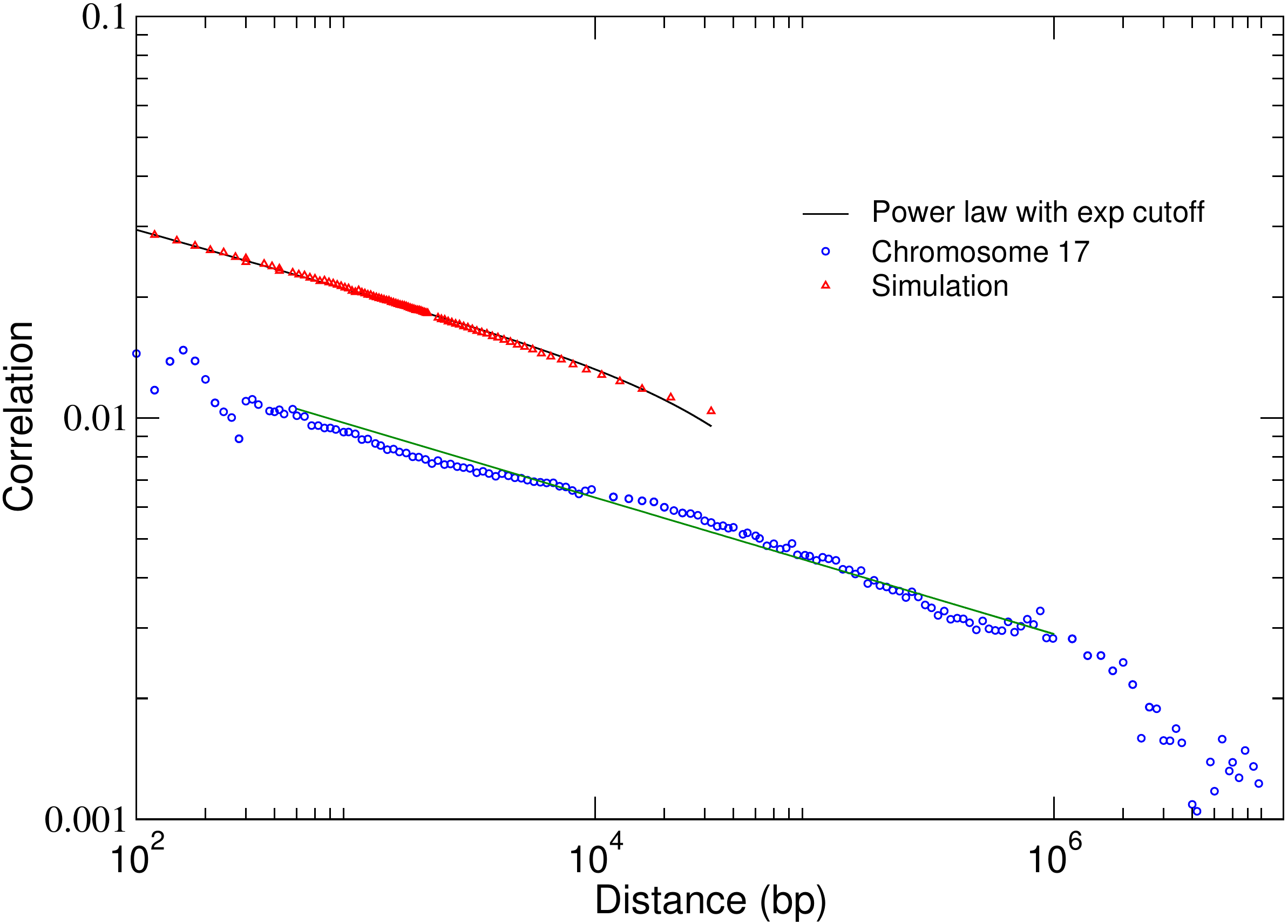}
  \end{center}
  \caption{\footnotesize{Color online. Comparison of the base-base correlations in the chromosome 17 (lower part of the figure, blue circles) 
  with the result of a simulation (upper part of the figure, red triangles) of the long range Ising
  model with $\alpha=1.75$, $h=0$ and $\beta>\beta_c$. The mean magnetisation is $\sim 5\%$ in both cases.}} \label{pic:chrom17}
\end{figure}

All the above considerations descend from well known results on the 1d long range Ising model.
What is probably less known is that, (contrary to the intuition we have from the short range Ising models), this model, due to the peculiar long range interaction term, 
can sustain long range correlations in a very robust way without the need of fine tuning the value of the two relevant coupling constants ($\beta$ and $h$) to the 
critical value. 
This is due to the fact that as we leave the critical point the correlation length decreases much more slowly than in the short range model leaving a large window 
within which the spin-spin correlator decreases with the same power law of the critical point. 
For instance in the short range 2d Ising model, for values of $h$ such that the magnetisation is of the order of $5\%$  the correlation length would be of few 
lattice spacings while in the long range Ising model it reaches $10^5$ lattice spacings.
To support this observation we performed extensive simulations 
for $h\not=0$ and $\beta \not= \beta_c$ using the very efficient  cluster  
algorithm of \cite{Luijten02}. 
We report as an example in fig. \ref{pic:chrom17} the result of a simulation performed at $h=0$ and $\alpha=1.75$ in the broken symmetric phase of the model, 
choosing the value of $\beta$ so as to have a mean magnetisation of $5\%$ (red triangles in the figure). We see that the correlation length is of the order of $2 \times 10^5$ bp 
and that in the range $10-10^5$ a power law behaviour with a value of the exponent (which we extracted from the simulations using exactly the same protocol 
which we used for the real DNA sequences) $\gamma\sim 0.16$ which is only slightly smaller than the one $\gamma=2-\alpha=0.25$ predicted (and observed) at the
critical point. We plot in the same figure for comparison the correlator of the chromosome 17 (which was chosen only because it has a value of $\gamma$ similar to the one
obtained in the simulation). The main lesson that we learn from these simulations is that, due to the long range correlators in the Hamiltonian, the model is very robust,
i.e. it is characterised
by a large scaling region with correlation lengths which, even for values of the magnetisation similar to the ones observed in the real sequences, reach hundreds of
kilobases and with values of $\gamma$ slightly smaller than the critical ones, but of the same order of magnitude.  

\section{Biological models}
As we mentioned in the introduction, in the past 20 years there were several attempts to understand long range correlations by linking them to other biological or physical
characteristics of the DNA chain. These proposals range from the association to the bending \cite{Goodsell94} or elastic \cite{Vaillant03} properties of the DNA chain 
to the hierarchical organization of chromosomes
\cite{Audit01,Audit02}, to the thermodynamic properties of DNA loops \cite{Vaillant05},
 to the role of folding and more generally of the 3d structure of 
chromatin  \cite{Grosberg93} or of Transposable Elements insertions in the E-R type of models \cite{messer_solvable_2005,li_study_1997}.  
Among these proposals these last two are particularly interesting for us due to the non-local nature of chromatin contacts and transposons insertions.  
Thanks to the recent advances in the study of the 3d structure of chromatin and of transposons classification \cite{RepeatMasker} we are now in the position to test them in a more
quantitative way.

\subsection{Chromatin contacts}
A possible role of chromatin contacts to explain  DNA correlations was proposed very early \cite{Grosberg93}, but was then  somehow abandoned when it was
realized that the scaling exponents of DNA correlations were much smaller than the typical scaling exponents of chromatin contacts that random polymer models would suggest. In
the past few years thanks to the remarkable advances in 3C (Chromosome Conformation
Capture) and Hi-C technologies \cite{Lieberman-Aiden09,nagano_single-cell_2013} these chromatin contact exponents were observed experimentally and the gap with the DNA
correlation scaling exponents was indeed confirmed. However we have seen above that the 1d Ising model could nicely explain, thanks to eq.(\ref{rel}) this gap. This prompted us to
reanalyze in more detail this proposal.

Hi-C technologies allow to obtain detailed genome-wide information on the physical contacts 
among distant genomic regions. The idea emerging from these experiments
is that chromatin has a complex, `scale free' like organisation across a range
of spatial scales. The most impressive results of these studies has been the discovery of the so called Topological Associated Domains (TAD) \cite{dixon12,Nora12} which are
domains characterised by enriched levels of DNA-DNA contacts with an average contact probability,
$P_c(s)$, which decreases as a function of the genomic separation
approximately as a power law, $P_c(s)\sim s^{-\alpha_{TAD}} $, in the 0.5 - 7 Mb range.
The values of $\alpha_{TAD}$ depend rather strongly on the cell line and condition, ranging for instance 
from $\alpha_{TAD}\sim 1.6$ for embryonic stem cells to $\alpha_{TAD}\sim 1.1$ for lymphoblastoid cells in the interphase (see \cite{Barbieri12} for a review).
Several models have been proposed to describe this behaviour. Among the others an interesting proposal is the Strings and Binder Switch (SBS) model \cite{Barbieri12} which describes the
chromosomes as self-avoiding polymer chains with binding sites for diffusing molecules which mediate the DNA-DNA interactions. 
In order to create such an interaction  the two portions of the chromosome should share the same binding sequence. The evolutionary process which led to the formation of
these pairs of similar binding sequences is very similar to the one we discussed above. In order to create a contact the two binding sequences should evolve so as to become
similar. The probability of this event to occur decreases as a function of the distance along the DNA chain following a power law exactly as in eq.(\ref{prob}). Moreover
depending on the CG content of the binding sequence we may have an overall drift modelled by the $h$ term in eq.(\ref{prob}).
This identification is appealing, but it faces at least three problems: 
\begin{itemize}
\item
The TAD exponents show a cell line variability larger than the scaling exponents of the DNA
correlations.
\item
The TAD exponents are in general slightly smaller than what would be needed to explain DNA correlations.
\item
The renormalization analysis discussed in sect. 2.3 suggests that the typical size of the sequences driving the correlations should be of the order of 100 bp, 
much larger than
the typical size of a transcription factor binding sequence.
\end{itemize}
There are a few possible directions which one could follow to address  these issues. Let us briefly discuss them: 

\begin{itemize}
\item
TADs are likely to be the results of the action of different families of binding proteins. In each cell line only a subset of
them is expected to be present, leading to a wide variety of TAD scaling exponents, while in the DNA sequence we see the signature of the union of all the corresponding
binding sequences.   
\item
Our numerical experiment (see sect.3) shows that as we leave the critical point we measure an "effective" scaling exponent slightly smaller than the one predicted by eq.(\ref{rel}).
This could explain the gap between the observed values of $\gamma$ and those obtained from  $\gamma=2-\alpha_{TAD}$
\item
It is well known that a few specific familes of  Transposable Elements (TE) (we discuss in detail TEs in the next section)  can bind in a specific way some transcription factors \cite{Jordan03} 
and are one of the tools the cell uses to convey genome-wide
combinatorial regulation \cite{Testori12}. If this holds true also for the proteins mediating the DNA-DNA interactions then 
we could identify their binding sequences with specific families of  Transposable Elements whose typical size is exactly the one predicted by the renormalization analysis. 
\end{itemize}

Overall we can say that a link between TAD organization and DNA correlations is plausible but that for the moment we are far to prove it. It is likely that future progress in understanding 3d chromatin contacts will also shed some light on this problem. 

Before ending this section let us stress that direct chromatin contacts is not the only way one can use Hi-C results to explain DNA correlations.
Another mechanism which could also  lead to long range sequence similarities
is the colocalization of coregulated genes (see for instance the recent study  
in the case of the human chromosome 19 in \cite{DiStefano2013}). This coregulation requires the presence of 
common regulatory sequences which, similarly to their target genes, should colocalize along the chromosome.
Also in this case however, in order to substantiate this hypothesis one should identify these regulatory sequences and test their power law behaviour.

The important observation for the scope of the present paper is that in both cases we expect that, if such a power decay exists, it should be driven by $\alpha_{TAD}$ which
is much larger than the exponent $\gamma$ of the DNA correlators. Our model, and in particular the relation $\gamma =2-\alpha$ of eq.(\ref{rel}) offers a nice explanation of this gap.

\subsection{Retrotransposon insertion as a possible source of long range correlations.}

Transposable Elements, thanks to their repetitive sequence and their widespread presence in the genome are natural candidates as putative drivers of  
long range DNA correlations. Moreover it was reported in  \cite{Repeat1,Repeat2} that the probability distribution $P(s)$ of the 
inter-transposon distance $s$ follows a power law behaviour $P(s)\sim s^{-\mu}$ . We devote this section to a critical exam of this issue. 

Transposable Elements (transposons) represent almost 45\% of the human genome (for a review see for instance \cite{Cordaux09}). 
They are genetic elements which are able to duplicate 
themselves and insert in a (almost) random way in the hosting genome. They strongly contributed to shape the genome of all higher eukaryotes.
Among the different transposon families a special role is played in the human case by the Alu and the LINE families which comprise the 10\% and 20\% 
respectively of the human
genome. LINEs are AT rich autonomous retrotranspons. Their full length  is of about 3kb, but most of their copies in the genome 
are truncated and thus the mean length is only of 430 bp. 
They are very successful transposons and are present in almost $1.5 \times 10^6$ copies in the
human genome.  Alus are non autonomous elements which are retrotransposed by the 
LINE machinery, they are about 300 bp long, CG rich and are probably the most successful non autonomous transposons in the primate 
lineage with more than $10^6$ copies in the human genome.   

There are indeed a couple of features of Transposable Elements that  support their possible role  in driving DNA correlations:
\begin{description}

\item{(i)}
The renormalization analysis discussed in sect. 2.3 suggests that the typical size of the sequences driving the correlations should be of the order of 100 - 1000 bps, 
which is exactly the typical size of most of the existing Transposable Elements in the genome.

\item{(ii)} Transposon insertion can be described using a non local Markov process with some analogies with the one that we proposed in sect.3. 
If we identify the first "spin" as one particular transposon,  
a new insertion of the same transposon  would represent a "successful" step of the Markov process in which the second spin is aligned with the
first one with  an excess probability with respect to a random choice. There are however two main differences with respect to our model. First, 
while in the transposon case we have 
the insertion of a new spin and the total length of the sequences increases, in our case we substitute an existing spin with a new one and the total length is fixed. 
Second, 
while transposon insertion is certainly a non-local process, there is no obvious reason (except for the observation of \cite{Repeat1,Repeat2} which however is limited to the 
short scale behaviour of inter-transposon separation) to assume that the insertion probability should follow a power law as required in our model.

\end{description}

\begin{figure}[h] 
 \begin{center}
  \includegraphics[width= 1.2\textwidth]{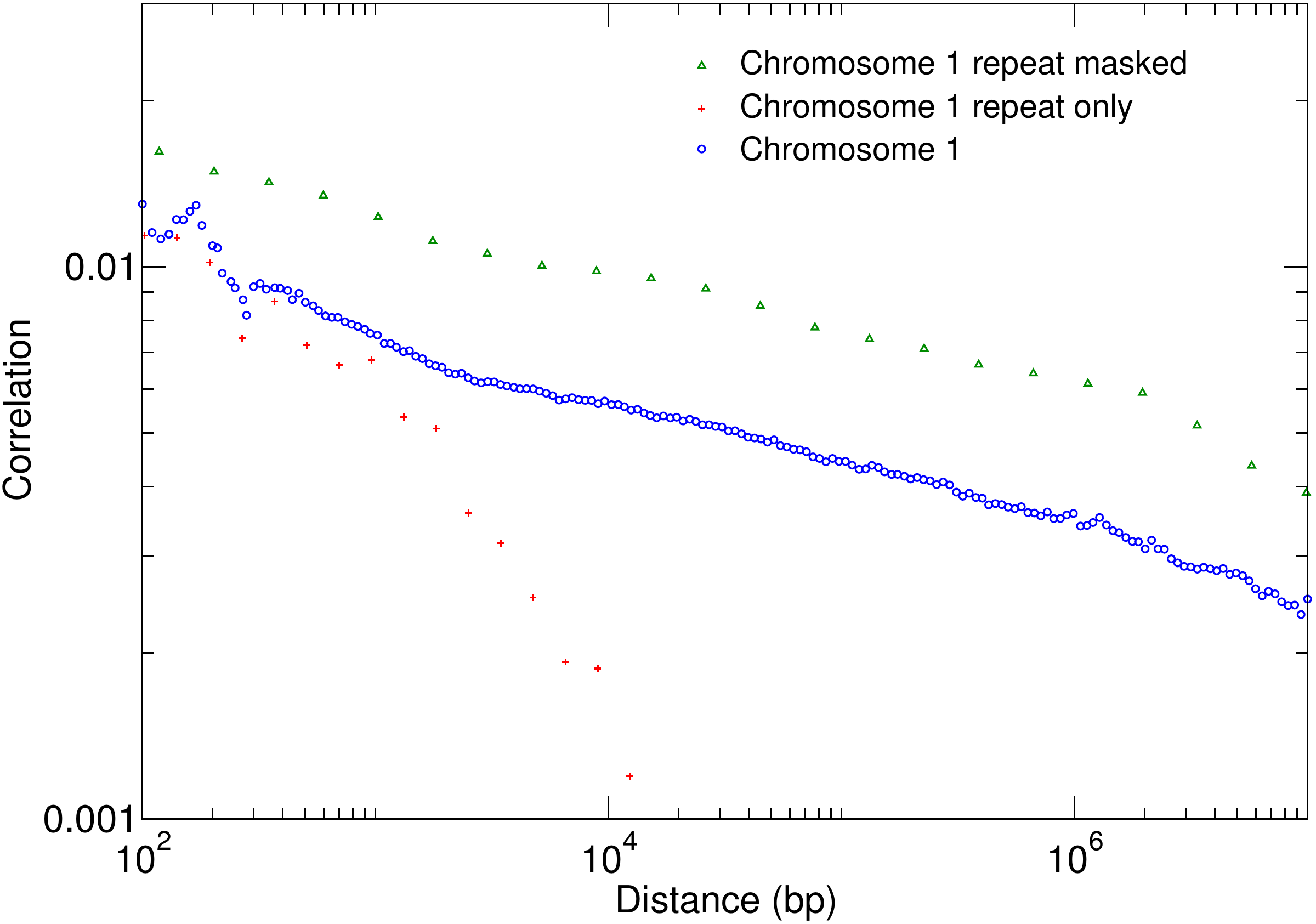}
  \end{center}
  \caption{\footnotesize{Color online. DNA correlations in chromosome 1 (blue circles) and in two artificial sequences obtained by eliminating transposons form the chromosome (green triangles), and by keeping only the transposons (red pluses).
  It is clear that DNA correlations are not affected by the elimination of transposons, while the artificial chromosome made of only
Transposable Elements shows no evidence of a long range correlation beyond 10kb.}} \label{newfig}
\end{figure}

To test the role of transposons we performed the same correlation analysis discussed above 
in two artificial sequences obtained (using the RepeatMasker software \cite{RepeatMasker}) first  by removing all the
transposons (which were substituted with zero in the sequence thus giving no contribution to the correlators)
and, second, by keeping only the trasposons and substituting with zero the rest of the sequence. The results of the analysis are reported in fig.\ref{newfig}. 

Looking at the figure it is easy to see that the correlations survive almost unchanged when the transposons are removed and that the artificial chromosome made of only
Transposable Elements shows no evidence of a long range correlation beyond 10kb, which is the typical scale of nearest neighbour insertions. Together these two observations
seem to exlude a role of transposons insertions in generating long range correlations. However it is important to stress 
that the above test involved the entire population of
trasposable elements as a whole. In view of the discussion of the previous section we cannot exclude that specific subfamilies of
transposons could instead play a role both in the formation of TADs and of DNA correlations.

\section{Concluding remarks and open issues.}

In this paper we have shown that long range DNA correlations can be described rather well by the 1d long range Ising model and that this description is rather robust, i.e.
it does not need a specific fine tuning of the model parameters. 
We have also shown that a simple evolutionary model in which distant portions of DNA
tend to have identical DNA sequences  can be mapped to a Markov model which has the 1d long range Ising model as stationary point.

There are a few open issues which we think should require further studies:
\begin{itemize}
\item
As we mentioned in the introduction, the optimal description of the long range nucleotide correlations could probably be achieved by a suitable combination of the 
Expansion-Randomisation model with our 1d Ising proposal. To this end it would be important to evaluate the changes in the duplication and insertion rates with time and across the
different species. Thanks to the increasing amount of sequencing data it is likely that precise measures of these rates will soon be available 
and will allow to tune the interplay between the two models.    
\item
It would be interesting to extend the present analysis to other organisms. Thanks to NGS studies we have now a rather precise knowledge of the retrotransposon repertoire
of several organisms\cite{Kofler12,Fiston14} and this could allow to test our results in a more extensive way.
\item 
It would also be important to further address the role of single families of Transposable Elements in this context. 
While the tests discussed in sect 4.2 exclude a role of repeated elements as a
whole, one cannot exlude a possible role of  selected subsets of transposons in shaping DNA correlations.
\end{itemize}

It is likely that in the near future,
thanks to the ongoing experimental efforts both on the sequencing side and on the reconstruction of the 3d structure of interphase chromosomes, much more data will be
available to address these issues and to deepen our understanding of the physical and evolutionary constraints which shaped the genomes of higher eukaryotes.   

\vskip 0.5cm

{\bf Acknowledgements:}
We thank  E. Domany, C. Micheletti, M. Osella and A. Rosa for useful discussions and suggestions.
This work was partially funded by the FSP grant GeneRNet.

\end{document}